# Imaging biofilms in three dimensions: modalities, quantitative readouts, and the path to four-dimensional measurement

Seongsoo Lee[1,2,3] and YongKeun Park[4,5,6*]
[1]Aging Research Group, Honam Regional Center, Korea Basic Science Institute (KBSI), Gwangju, Republic of Korea
[2]Department of Systems Biotechnology, Chung-Ang University, Anseong, Republic of Korea
[3]Department of Life Science, Hanyang University, Seoul, Republic of Korea
[4]Department of Physics, Korea Advanced Institute of Science and Technology (KAIST), Daejeon, Republic of Korea
[5]KAIST Institute for Health Science and Technology, KAIST, Daejeon, Republic of Korea
[6]Tomocube Inc., Daejeon, Republic of Korea
[*] Correspondence: yk.park@kaist.ac.kr (Y.K.P.)

### Abstract

Biofilms are spatially structured microbial communities whose architecture, chemistry, mechanics, and cellular states evolve over time. Bulk assays and two-dimensional projections remain useful, but cannot alone resolve how these properties vary with depth or change during growth, treatment, dispersal, and regrowth. Imaging provides complementary routes to three-dimensional measurement: fluorescence microscopy supplies molecular, taxonomic, and functional specificity; optical coherence tomography resolves mesoscale architecture and dynamics; quantitative phase imaging and holotomography report refractive index and biomass-related changes; Raman methods provide chemical and metabolic contrast; and Brillouin microscopy probes mechanical response. We compare these modalities using four independent descriptors—contrast provenance, live volumetric capability, perturbation, and demonstrated biofilm use—and connect their signals to quantitative biological readouts. No single modality simultaneously maximizes spatial coverage, resolution, acquisition speed, molecular specificity, and low perturbation. Implementations from any contrast class can serve as a longitudinal backbone when perturbation is empirically controlled at the relevant spatial and temporal scale, while molecularly specific measurements remain indispensable for identifying species, molecules, and functional states. We therefore frame four-dimensional biofilm measurement as a validated measurement architecture that integrates a low-perturbation volumetric backbone with spatially registered, molecularly specific measurements acquired continuously or at predefined validation points. Achieving this integration will require compatible cultivation formats, controlled imaging dose, shared quantitative parameters, and robust cross-modality registration.

### Introduction

Biofilms are three-dimensional (3D), gradient-stratified, dynamically remodeling microbial communities embedded in a self-produced extracellular matrix (Costerton et al. 1995; Hall-Stoodley et al. 2004; Flemming et al. 2016; Rumbaugh and Sauer 2020). Their architecture controls transport, cellular state, matrix organization, interspecies interactions, and responses to perturbation across clinical, engineered, and environmental settings, although estimates of disease prevalence remain context-dependent (Costerton et al. 1999; De La Fuente-Nunez et al. 2023); antimicrobial-resistance priorities include organisms for which biofilm-mediated persistence contributes to outcome (Sati et al. 2025); and the consequences of biofilm growth extend across medicine, industry, and the environment (Cámara et al. 2022). These problems share a measurement requirement: spatial organization and temporal change must be retained rather than collapsed into a bulk endpoint.

Many widely used assays—including broth minimum inhibitory concentration, time-kill curves, colony-forming-unit enumeration, and two-dimensional crystal-violet assays—reduce a structured community to population-averaged growth, survival, or biomass. These measurements remain valuable for screening and comparison, but they do not determine where molecules accumulate, which cells respond at depth, how the matrix changes, or whether an apparently disrupted community later disperses or regrows. Antimicrobial response therefore provides a demanding test case for 3D imaging; the expanding AMP pipeline is one example in which efficacy depends on spatially and temporally distributed penetration, cellular response, matrix remodeling, and regrowth (Magana et al. 2020; Ju et al. 2025). Yet *in vitro* AMP efficacy in planktonic cultures rarely transfers to biofilm-bearing models (Mookherjee et al. 2020). This mismatch illustrates the broader principle that 3D imaging is not merely a more detailed visualization but a way to connect spatial organization with biological function.

This review examines imaging methods for connecting biofilm organization with biological function and the quantitative readouts that can be extracted from them. We define four-dimensional (4D) measurement as repeated, spatially registered acquisition of 3D information from the same biofilm over time. The modalities considered here provide complementary contrasts rather than a hierarchy: fluorescence, affinity probes, and genetically encoded reporters supply molecular, taxonomic, and functional specificity; optical coherence tomography, quantitative phase imaging and holotomography, Raman imaging, and Brillouin microscopy provide structural, biomass-related, chemical, metabolic, or mechanical information with different acquisition and perturbation profiles. We classify implementations separately by contrast provenance, live volumetric capability, perturbation, and demonstrated biofilm use because instrument capability does not itself establish biological validity. Antimicrobial action—penetration, depth-resolved cellular response, matrix remodeling, and dispersal or regrowth—serves as a demanding test case rather than the exclusive subject. Our central argument is that 4D biofilm imaging should be designed as a validated measurement architecture: a longitudinal 3D channel selected for low perturbation at the relevant scale, integrated with molecularly specific measurements acquired continuously or at predefined validation points.

**From 3D architecture to 4D biofilm dynamics**

Biofilms are 3D structures whose development and responses are intrinsically time-dependent. 4D imaging captures this evolution through repeated, spatially registered volumetric measurement, whether the biofilm develops spontaneously or responds to an external perturbation (Figure 1a). The defining structural elements of a mature biofilm — micro-colonies, mushroom-shaped towers, water channels, and the surrounding extracellular polymeric substance (EPS) matrix — are organized across tens to hundreds of micrometres of depth (Costerton et al. 1995; Hall-Stoodley et al. 2004). Within this volume, oxygen, nutrients, signaling molecules, and applied antimicrobials exhibit depth-dependent gradients spanning one to two orders of magnitude, producing phenotypically distinct subpopulations — active, dormant, persister — that are spatially structured rather than genetically distinct (Stewart and Franklin 2008; Lewis 2010; Brauner et al. 2016). A two-dimensional projection collapses this geometry to an areal density; plate-based assays cannot distinguish whether a treatment removes biomass, opens channels, or detaches an intact layer over a viable residue. And because formation, maturation, dispersal, and reassembly form a continuous life cycle spanning minutes to days (Rumbaugh and Sauer 2020; Sauer et al. 2022), endpoint imaging captures architecture but loses the trajectory. The same biofilm volume three days after treatment can reflect productive killing, transient remodeling before dispersal, or post-dispersal regrowth — scenarios with opposite clinical implications.

Antimicrobial challenge provides a demanding 4D test case rather than the sole organizing principle for biofilm imaging. It resolves into four linked questions: where the agent goes, where cells respond or die, how the matrix changes, and whether the community disperses or regrows. Penetration limits, depth-stratified tolerance, matrix sequestration, and post-treatment regrowth are spatially and temporally distributed processes that bulk assays cannot localize. Host immunity, flow, nutrient delivery, polymicrobial interaction, and pharmacokinetics modify each link. Three of the four questions require longitudinal measurement, while several require chemical or biological specificity that structural contrast alone cannot provide. The central design problem is therefore to combine a low-perturbation longitudinal backbone with targeted molecular and functional validation without breaking continuity of observation (Figs. 1b–c; Box 1).

***Where does the agent go? (penetration).*** Whether and how fast an antimicrobial diffuses through the EPS to reach viable cells at depth is the foundational question, and the one least served by bulk concentration. Small uncharged molecules diffuse through hydrated EPS at 60–80% of their free-solution rate, but cationic AMPs and aminoglycosides are sequestered by anionic polysaccharides and extracellular DNA, limiting effective penetration to a fraction of biofilm thickness (Walters et al. 2003; Stewart 2003; Tseng et al. 2013). Direct gradient measurements remain rare and are endpoint snapshots — 3D ToF-SIMS on *ex vivo P. aeruginosa* lung biofilm (Davies et al. 2017), hyperspectral SRS of alkyne-tagged vancomycin in *S. aureus* biofilm (Bae et al. 2019). Direct agent-distribution measurements require a chemically specific channel. The unresolved need is spatial registration of a validated agent-specific signal with a low-perturbation longitudinal record of cellular and matrix response in the same mature biofilm (Stewart 2025).

***Where does it kill? (depth-resolved viability).*** Where cells die — at the surface, in the active periphery, or in the anoxic core — distinguishes a productive antimicrobial from a cosmetic one. Bulk CFU averages these compartments; confocal Live/Dead recovers the spatial information but inherits photobleaching and dye artifacts. The survivors are *tolerant* — a transient, non-heritable capacity distinct from genetic resistance (Lewis 2010; Brauner et al. 2016) whose spatial structure bulk endpoints discard. Light-sheet microscopy has linked translation inhibition to architectural collapse in *V. cholerae* biofilm (Díaz-Pascual et al. 2019), and $D_2O$-SRS has shown phenazine-driven metabolic heterogeneity underlying *P. aeruginosa* tolerance to several antibiotics but not colistin (Schiessl et al. 2019). These methods report complementary aspects of architecture, translation, or metabolism; neither is a universal measure of death. The unresolved requirement is a depth-resolved cellular-response readout validated against an orthogonal viability or activity assay within the same experiment.

***What does it do to the matrix? (EPS remodeling).*** The matrix is a dynamic, actively remodeled compartment — it shrinks, swells, dissolves, and reorganizes within minutes of challenge, and its trajectory predicts collapse, dispersal, or regrowth. Comprising ~90% of biofilm dry mass, the *matrixome* serves drug sequestration, enzyme retention, and mechanical cohesion among other functions (Flemming and Wingender 2010; Flemming et al. 2016; Karygianni et al. 2020). Imaging it has depended on lectin staining (partial coverage), electron microscopy (~60% dehydration collapse (Relucenti et al. 2021)), or single-timepoint mass spectrometry (Van Hoogstraten et al. 2024). Dynamic matrix readouts include refractive index (RI) mapping by holotomography (Buzalewicz et al. 2024), mechanical contrast by Brillouin microscopy (Karampatzakis et al. 2017); composition-specific fluorescence and vibrational imaging provide complementary identity, but no single contrast supplies matrix identity, architecture, and mechanics simultaneously.

***Does the community reassemble? (dispersal and regrowth).*** A regimen that eliminates 99% of a biofilm but lets the remaining 1% disperse and reassemble is a clinical failure dressed as a laboratory success. Dispersal is an active, regulated stage triggered by nitric oxide, glutamate, and *cis*-2-decenoic acid and mediated by matrix-degrading enzymes (Rumbaugh and Sauer 2020; Bridges et al. 2020), and the dispersed phenotype is often hyper-virulent. Light-sheet imaging has resolved dispersal trajectories at single-cell resolution (Qin et al. 2020; Bridges et al. 2020), but these records depend on genetic tractability and photon-dose budgets that limit observation to model species over hours-to-days. Long-duration volumetric tracking across polymicrobial communities and clinical isolates remains limited irrespective of contrast class. The target experiment requires a low-dose, low-perturbation longitudinal backbone coupled to scheduled identity and viability validation.

### A framework for comparing biofilm imaging modalities

Imaging modalities can be compared using four independent descriptors. Contrast provenance distinguishes intrinsic, tracer-assisted, and exogenous-label implementations. Live volumetric capability separates repeated 3D acquisition from surface, projection, or terminal measurements. Perturbation includes probe chemistry, genetic modification, mechanical contact, acquisition dose, and destructive preparation. Demonstrated biofilm use separates instrument capability from biological validation. Table 1 reports these dimensions independently; Figure 2 maps resolution and acquisition speed without ranking modalities; and Figure 3 illustrates representative structural, molecular, mechanical, and longitudinal readouts. A complementary recent review examines spatiotemporal dynamics from a biofilm-development perspective (Espinoza Miranda et al. 2025). Here we emphasize the connection between modality-specific signals, quantitative readouts, and longitudinal experimental design.

***Fluorescence-based volumetric microscopy.*** Confocal microscopy provides probe-defined species, matrix, and cell-state channels and supports the field's most mature quantitative pipelines (Mhade and Kaushik 2023), while usable depth and duration depend on probe distribution, scattering, bleaching, and cumulative dose. Light-sheet microscopy reduces out-of-focus exposure and tracks single-cell trajectories across days, resolving architectural transitions, collective flow, and antibiotic-induced collapse (Drescher et al. 2016; Díaz-Pascual et al. 2019; Qin et al. 2020). Multiphoton and fluorescence-lifetime implementations extend imaging depth or provide endogenous metabolic contrast (Gries et al. 2020). Across these implementations, quantitative outputs include intensity and lifetime fields, segmented biovolume and matrix distributions, and cell trajectories; their biological specificity depends on validation of the probe, reporter, or endogenous lifetime assignment. These platforms may serve as the longitudinal backbone when reporter stability, optical access, and dose permit, or as molecularly specific validation channels in a multimodal workflow. Other volumetric approaches provide complementary structural, chemical, and mechanical readouts and are considered below. Bioluminescence imaging is discussed with translational scale bridging (Avci et al. 2018).

***Contrast provenance and perturbation.*** Contrast provenance has three classes. Intrinsic implementations derive signal from the specimen without a contrast-generating reagent (for example, RI, backscatter, acoustic phonons, force–distance, electron scattering, or endogenous molecular vibration). Tracer-assisted implementations introduce an isotope, chemical tag, or radiotracer; exogenous-label implementations use fluorescent dyes, affinity probes, viability stains, or genetically encoded reporters. These classes describe where contrast originates; they do not rank perturbation. Perturbation must be evaluated separately from probe chemistry, genetic modification, mechanical contact, acquisition dose, and destructive preparation. The $D_2O$- and alkyne-based SRS applications discussed here are tracer-assisted rather than intrinsic. For cationic AMPs, fluorescent conjugation systematically alters amphipathicity, membrane partitioning, and kill kinetics (Choi et al. 2016) [Box 1]; lectin probes likewise capture only a fraction of the polysaccharide repertoire (Reichhardt and Parsek 2019). The appropriate reference depends on the biological claim: intrinsic measurements preserve continuity without a contrast-generating reagent, whereas tracer-assisted or exogenous-label measurements provide molecular identity. Neither class is universally non-perturbative or a default standard.

**Representative longitudinal backbones and complementary readouts**

Longitudinal approaches should be selected by biological scale, imaging depth and field of view, required specificity, acquisition rate, and dose budget—not by contrast provenance alone. Confocal, light-sheet, and multiphoton microscopy can provide molecularly resolved volumetric trajectories; OCT provides mesoscale structural continuity; quantitative phase imaging (QPI), including holotomography (HT), reports refractive-index and biomass-related dynamics; SRS supplies chemical or metabolic contrast; and Brillouin microscopy adds mechanical information. AFM provides interface-resolved mechanics rather than internal volumetric coverage. Figure 3 illustrates different measurement roles rather than identifying a preferred platform.

***Optical coherence tomography*** provides a mesoscale structural view—millimetre field of view, 1–3 mm depth, and ~10 µm axial resolution. It resolves architecture, deformation, detachment, and regrowth at clinically relevant scales, but without an agent-specific contrast channel it cannot directly report antimicrobial concentration, penetration, or depth-resolved viability. It reconstructs depth-resolved backscatter by low-coherence interferometry and can extend imaging depth beyond confocal in scattering biofilms, at coarser spatial resolution. The Wagner–Horn group formalized the OCT-biofilm parameters (thickness, roughness, porosity, biovolume) (Wagner and Horn 2017), with extensions to viscoelastic deformation under hydrodynamic load (Figure 3e) (Valladares Linares et al. 2016), membrane-bioreactor biofouling, catheter-based imaging of endotracheal-tube biofilm in ICU patients (Dsouza et al. 2019), and OCT–bioluminescence fusion (Demidov et al. 2025). Its quantitative outputs are primarily structural—thickness, roughness, porosity, biovolume, deformation, and detachment—whereas backscatter intensity is protocol-dependent and does not by itself establish agent identity or cellular viability.

***Stimulated Raman scattering and related vibrational imaging*** span intrinsic vibrational contrast and tracer-assisted chemical imaging. In the cited biofilm implementations, $D_2O$ and alkyne-tagged drugs make the measurements tracer-assisted rather than intrinsic, although they avoid exogenous fluorophores. These implementations provide chemical or metabolic specificity at reported depths of approximately 50–100 µm, with achievable depth and volume rate depending on spectral acquisition and specimen. Hyperspectral SRS with alkyne-tagged vancomycin quantified drug diffusion and intracellular accumulation in live *S. aureus* biofilm (Bae et al. 2019); $D_2O$-SRS linked metabolic heterogeneity to antibiotic tolerance in *P. aeruginosa* (Schiessl et al. 2019). Related Raman implementations extend these roles: SERS has mapped pyocyanin gradients (Bodelón et al. 2016), while the hybrid SRS-FISH workflow adds taxonomic identity to metabolic measurements (Ge et al. 2022). These studies demonstrate agent-localization and metabolic-state measurements in biofilms, but repeated three-dimensional acquisition throughout a mature-biofilm antimicrobial challenge was not established in the cited examples.

***Holotomography*** is a 3D quantitative phase imaging (QPI) implementation that reconstructs the RI distribution of live specimens from multi-angle interferometric phase. It supports repeated three-dimensional acquisition without reliance on fluorophore stability; illumination dose, phase stability, angular sampling, scattering, and usable depth remain protocol- and specimen-dependent. RI measurements can support dry-mass- or concentration-related estimates when a composition-appropriate RI increment is assumed or calibrated (Barer 1953; Park et al. 2018; Kim et al. 2024). Single-bacterium studies have reported RI dynamics during antimicrobial exposure (Kim et al. 2017; Oh et al. 2020; Choi et al. 2021; Kim et al. 2022; Shin et al. 2023; Kim et al. 2023). Biofilm studies have quantified species- and time-dependent volume and reported a reproducible RI shift associated with cell state under the tested species, medium, and growth conditions (Buzalewicz et al. 2023; Buzalewicz et al. 2024). HT has also imaged antimicrobial-peptide-associated changes and early biofilm-development dynamics in multidrug-resistant *Acinetobacter baumannii*, with efficacy evaluated by conventional assays (Kumar et al. 2025). Thus, HT is one possible cellular-scale longitudinal structural backbone; it does not intrinsically identify the antimicrobial, species, matrix component, or universal viability state. Heterogeneous-EPS mass interpretation requires composition assumptions or correlative validation.

***Brillouin microscopy*** provides intrinsic mechanical contrast by probing thermal acoustic phonons in live 3D specimens. It measures a frequency shift related to longitudinal mechanical response through refractive index and density (Prevedel et al. 2019; Kabakova et al. 2024). Brillouin imaging of live *P. aeruginosa* biofilms has reported spatial gradients in Brillouin frequency shift, interpreted as differences in local mechanical state and hydration (Karampatzakis et al. 2017; Mattana et al. 2017). The shift is a mechanical contrast rather than a direct stiffness or absolute-modulus measurement: quantitative interpretation requires RI, density, and constitutive assumptions. Co-registered RI imaging can reduce one source of uncertainty, but no single modality supplies an assumption-free modulus map.

***Interface-resolved mechanical companion: in-liquid atomic force microscopy*** uses intrinsic force–distance and topographic contrast to measure nanometre-scale surface morphology, stiffness-related response, and adhesion at accessible biofilm–liquid interfaces (Beaussart et al. 2014; Dufrêne and Persat 2020). High-speed AFM tracked the kinetics of AMP CM15 on individual *E. coli* cells (Figure 3b) (Fantner et al. 2010); AFM combined with ATR-FTIR on a six-hour nascent *P. fluorescens* biofilm under dermaseptin challenge (Quilès et al. 2016). Recent interior microrheology of live *V. cholerae* biofilm (Ohmura et al. 2024) provides a distinct, complementary mechanical precedent for the biofilm interior. Repeated AFM maps add longitudinal information at accessible surfaces but do not constitute volumetric 4D imaging. The cited antimicrobial precedents concern individual cells or a nascent biofilm; a mature-biofilm internal-volume workflow is outside AFM's geometry.

***Terminal hydrated-state companion: environmental and atmospheric SEM*** provides intrinsic electron-scattering contrast from hydrated or uncoated biofilms at nanometre scale. Atmospheric SEM through a silicon-nitride membrane window maintains an aqueous specimen environment and has resolved water channels, intercellular nanotubes, membrane vesicles, eDNA fibrils, and curli filaments in MRSA and *E. coli* biofilm (Nishiyama et al. 2014; Sugimoto et al. 2016). These are beam-sensitive, single-time-point structural measurements rather than *live* longitudinal acquisition.

***Complementary terminal validation modalities.*** These modalities provide high-resolution structural or chemical endpoint measurements but do not preserve longitudinal continuity. FIB-SEM and cryo-electron tomography with cryo-FIB milling provide nanometre-scale ultrastructure and, in selected cryo-ET subvolumes, sub-5-nm molecular detail, including the *P. aeruginosa* CdrA adhesin mediating cell–cell junction in mature biofilm (Melia et al. 2021). Cryo-orbitrap SIMS and MALDI imaging map molecular composition and antibiotic gradients (F. Zhang et al. 2021; Shen et al. 2025); soft X-ray tomography reaches 50 nm isotropic resolution in cryo-vitrified *B. subtilis* biofilm (Chatzimpinou et al. 2025); 3D-STORM and micro-CT add super-resolved matrix architecture and diffusion-barrier mapping (Berk et al. 2012; Keren-Paz et al. 2018). These methods provide high-resolution structural or chemical reference measurements but generally report state rather than trajectory.

These modalities should be interpreted as components of a measurement architecture rather than as competitors for a single preferred position. Fluorescence-based methods offer the strongest molecular, taxonomic, and functional specificity; OCT, quantitative phase imaging, Raman, and Brillouin approaches offer complementary structural, chemical, and mechanical measurements across different scales; and interface or terminal methods provide orthogonal validation. No single modality simultaneously maximizes spatial coverage, resolution, speed, specificity, quantitative interpretability, and low perturbation. The practical question is therefore which combination measures the biological readout with sufficient continuity and validation.

**From modality-specific signals to quantitative biofilm readouts**

Raw 3D image stacks are not comparable biological measurements until they are reduced to defined parameters with known uncertainty. Fluorescence-confocal imaging has the most mature analysis ecosystem, beginning with COMSTAT (Heydorn et al. 2000) through PHLIP, daime, and deep-learning BCM3D (Daims et al. 2006; Mueller et al. 2006; M. Zhang et al. 2020) to BiofilmQ, now the de facto standard with sixty-plus parameters per voxel (Hartmann et al. 2021), with the Minimum Information About a Biofilm Experiment (MIABiE) guidelines codifying cross-laboratory reporting (Allkja et al. 2020). These fluorescence-derived pipelines provide the field's most mature quantitative framework, although probe distribution, depth attenuation, segmentation, and reporter biology must be included in uncertainty and quality control. OCT, HT, SRS, Brillouin microscopy, and AFM produce different quantitative fields—backscatter-derived architecture, refractive index, molecular vibration, mechanical response, and force–distance maps—that cannot be made comparable by applying one segmentation pipeline. The field needs both shared biological outputs, such as thickness, biovolume, depth-resolved response, matrix change, and regrowth, and modality-specific reporting of calibration, resolution, dose, uncertainty, and validation. BISCAP provides one OCT-specific example (Narciso et al. 2024).

**Molecular specificity, perturbation, and validation**

Fluorescent conjugation provides direct agent localization, but for cationic antimicrobial peptides it can also alter the molecule being measured. AMP membrane activity depends sensitively on cationic–hydrophobic balance; appending a fluorophore (TAMRA, Cy3, FITC, or BODIPY) adds bulk, hydrophobicity, and sometimes net charge. Reported conjugates of magainin, LL-37, melittin, and dermaseptin show MIC shifts, altered oligomerization, or changed lysis kinetics (Choi et al. 2016).

Each contrast class has a distinct failure mode. Intrinsic measurements preserve continuity without a contrast-generating reagent but may lack biological or chemical identity. Tracer-assisted and exogenous-label measurements can identify the agent, species, matrix component, or cellular state, but their perturbation and dose must be validated for the specific experiment. Low-dose fluorescence can itself provide the longitudinal backbone when reporters are stable and biologically appropriate; intrinsic imaging is not the default reference in every setting. For antimicrobial conjugates, the labeled and parent compounds should be compared directly when the conjugate is used as a proxy. The design principle is to choose the least perturbative combination that still measures the required biological identity; no contrast class is ground truth by default.

## Imaging beyond model biofilms

The modality framework is most easily established in monospecies biofilms grown on coverslips or in flow cells, but its value is tested in infected tissue, polymicrobial communities, and engineered environmental systems. These settings alter optical access, scattering, field of view, species complexity, cultivation geometry, and the feasibility of repeated acquisition. They therefore change which modality can serve as a longitudinal backbone and which molecular validation is required, rather than uniformly favoring either labeled or intrinsic contrast.

***Clinical and native-tissue biofilms.*** In the patient the biofilm is embedded in tissue rather than grown on glass, and imaging now measures the gap between the two directly. In cystic fibrosis airways, peptide–nucleic-acid fluorescence in situ hybridization (FISH) with matrix immunolabeling resolves mucoid Pseudomonas aeruginosa as matrix-embedded aggregates suspended in conductive-zone mucus that persist through intensive antibiotic therapy (Bjarnsholt et al. 2009). Hydrogel tissue-clearing combined with FISH quantifies aggregate biovolume and matrix directly in patient sputum(Jackson et al. 2020), and paired comparison shows that laboratory isolates fail to reproduce the in situ aggregation state — a measurement-level statement of the same bench-to-bedside discordance this review addresses (Morris et al. 2024). The aggregated architecture survives even the most effective modern therapy: CFTR-modulator treatment lowers airway bacterial burden yet leaves roughly a fifth of P. aeruginosa persisting as large aggregates (Morris et al. 2026), a clinical-scale readout of the dispersal-and-regrowth question. That result is the clinical counterpart of the chlorine finding discussed below. Two literatures that share no clinical setting, no agent, and no endpoint converge on the same discordance, which is what makes it a property of the measurement rather than of any one model. The pattern recurs in chronic wounds, where FISH and confocal microscopy localize P. aeruginosa and Staphylococcus aureus as tissue aggregates surrounded by a pathogen-specific but ineffective neutrophil response (Fazli et al. 2011), and comparable tissue-embedded aggregates are recovered from implant-associated pseudarthrosis, where imaging separates chronic infection from sterile mechanical failure (Ohrt-Nissen et al. 2019). These are fixed-tissue snapshots; none follows an antimicrobial encounter live. Two translational bridges frame the missing middle. At the bedside, point-of-care autofluorescence imaging detects biofilm-forming bacterial burden across hundreds of wounds that clinical inspection alone misses (Le et al. 2021); at whole-body scale, bacteria-specific molecular imaging — from technetium-99m-ubiquicidin to newer gallium-68-maltohexaose positron-emission tomography that discriminates infection from sterile inflammation — reports device-associated infection *in situ* (Wegener et al. 2026). Between the fixed-tissue snapshot and the whole-body scan lies a scale and validation gap: live volumetric imaging of the antimicrobial encounter coupled to agent- and cell-specific validation. Intravital microscopy begins to close it — confocal imaging in a murine model shows host immune cells physically excluded from, and slowed at, the S. aureus biofilm surface, imaging the penetration barrier to host defense *in vivo* (Sauvat et al. 2020) — yet it relies on fluorescent reporters and so inherits the label caveats set out above.

The clinical end of the pipeline operates at a different spatial scale. Bioluminescence imaging of luciferase-tagged bacteria provides longitudinal whole-mouse burden in implant, prosthetic-joint, and chronic-wound biofilm models and is the preclinical workhorse for AMP and phage efficacy (Bernthal et al. 2010; Wang et al. 2017). Translation to humans uses bacteria-specific positron emission tomography/single-photon emission computed tomography (PET/SPECT) probes: technetium-99m-labeled ubiquicidin 29–41 ($^{99m}$Tc-UBI) — itself a cationic AMP fragment — accumulates at infection and discriminates it from sterile inflammation, and is in routine clinical use for musculoskeletal infection (Welling et al. 2002; Sathekge et al. 2018; Ordonez et al. 2019). These scales are linked by calibrated biological endpoints rather than by a shared contrast mechanism. Microscopy resolves local structure and mechanism, bioluminescence follows whole-animal burden, and PET/SPECT localizes infection clinically. Translation requires prespecified correspondence among these outputs because none validates another automatically.

***Polymicrobial and host-associated communities.*** Real infections are seldom monospecies, and spatial arrangement — not composition alone — sets the antimicrobial outcome. Combinatorial-labeling spectral FISH resolves fifteen taxa simultaneously and revealed the micron-scale 'corncob' biogeography of human dental plaque, in which anaerobes occupy protected interiors and aerobes the periphery (Valm et al. 2011; Mark Welch et al. 2016), with individual microhabitats occupied by narrowly defined taxon subsets (Morillo-Lopez et al. 2022). In dental caries the same spatial logic is at once pathogenic and protective: intact biofilms show Streptococcus mutans as an inner core encased in a corona of other species, creating both localized acidic foci and a physical barrier that shields the community from antimicrobials (Kim et al. 2020) — a direct image of spatial structure conferring tolerance that no planktonic assay can anticipate. Chemical imaging sharpens the point: ratiometric confocal pH mapping resolves the acidic microhabitats within living cross-kingdom biofilms that govern both virulence and local drug efficacy (Schlafer and Frost Kristensen 2020). Function follows this architecture under treatment — in a P. aeruginosa–S. aureus wound-biofilm model, confocal imaging shows single-agent therapy shifting the two-species balance rather than clearing it, arguing that combination regimens be designed against imaged spatial co-localization (Rubio-Canalejas et al. 2022). The host interface adds a further compartment: live imaging in bladder-chip and organoid models resolves intracellular bacterial communities and wall-penetrating solitary cells that evade both antibiotics and neutrophil swarms (Sharma et al. 2021), spatially distinct refuges invisible to a bulk minimum inhibitory concentration. These studies demonstrate why molecular and taxonomic specificity is indispensable in polymicrobial and host-associated settings. Extending low-perturbation volumetric imaging into this regime will require it to be registered with, rather than substituted for, validated species- and host-specific channels.

***Environmental and industrial biofilms.*** The measurement problem is not confined to medicine, and the environmental setting is where longitudinal structural imaging is already well developed. Optical coherence tomography, a mesoscale structural modality, has become an established non-destructive probe of engineered biofilms: it resolves membrane-biofouling formation, internal void structure, and detachment in real time (Dreszer et al. 2014) and tracks how drinking-water biofilm thickness and structural response change over months of chlorine or chloramine exposure rather than simply eroding (Shen et al. 2016). The environmental literature also provides a useful test of this review's thesis: under chlorine, more than 95% of biofilm cells can enter a viable-but-non-culturable state while the physical architecture persists intact, and plate counts decrease while OCT shows that substantial biofilm architecture persists (Guo et al. 2023). The culture and structural readouts measure different quantities and can therefore diverge. Neither adjudicates biological outcome alone — culture under-reports survivors by construction, and persisting architecture does not by itself establish that those survivors remain consequential — but that is the point, because their combination makes the disagreement visible. The same non-destructive readout captures how combined chemical and hydraulic cleaning redistributes rather than removes biofilm in water-reuse systems (Lequette et al. 2021), and how non-chemical interventions such as germicidal ultraviolet light reshape biofilm density on membranes (Rho et al. 2022). In food processing, quantitative confocal structure parameters of Listeria monocytogenes biofilm correlate with benzalkonium-chloride tolerance, linking imaged architecture to disinfectant susceptibility on food-contact surfaces (Nogueira et al. 2023). One industrially critical domain remains incompletely examined longitudinally—microbially influenced corrosion, where biofilm–metal interfaces are still resolved largely by terminal electron microscopy. Progress will require structural continuity coupled to chemical and functional validation.

## Designing reproducible live 3D experiments

A modality's capability on paper is realized only through the culture format it images; for live biofilm imaging, the cultivation system and the microscope function as a single instrument. Dynamic flow produces more uniform, spatially organized, and reproducible architecture than static culture, so results from the two formats are not interchangeable, and microfluidic platforms now couple precisely controlled laminar flow to single-cell-resolution, label-free observation of adhesion, growth, and in situ antimicrobial killing (Straub et al. 2020). The classical parallel-plate flow cell remains the reference format whose cross-laboratory adoption underpins reproducible confocal biofilm imaging (Crusz et al. 2012); low-shear drip-flow reactors extend imaging to air–liquid-interface biofilms amenable to correlative chemical mapping (Li et al. 2018); and multichannel and standardized microtiter-format devices now link in situ imaging to species-resolved molecular readout and higher-throughput antimicrobial susceptibility testing (Cheah and Bae 2023; McLeod et al. 2023). Substrate is an equally active variable: implant-material surface roughness measured by atomic force microscopy spans nearly an order of magnitude between titanium alloy and bone cement, shaping the very architecture the microscope then records (Van Dun et al. 2023).

Long-duration live imaging requires an explicit dose budget. Phototoxicity and photobleaching cap how long a fluorescently labeled community can be followed — which is precisely why two developments matter for the dispersal-and-regrowth question: refractive-index-matched, hermetically sealed flow channels that permit multi-day lattice-light-sheet imaging with no observable photodamage (J. Zhang et al. 2021), and intrinsic-contrast modalities that avoid fluorophore bleaching while still requiring explicit illumination-dose control. The recurring lesson is that reproducibility in biofilm imaging is set upstream of the microscope — by flow, substrate, and growth condition — so that shared 3D quantification and reporting standards across modalities, discussed next, will deliver cross-laboratory comparability only if paired with standardized cultivation and reporting.

**Integrating modalities toward 4D measurement**

The longitudinal backbone should be selected according to biological process, spatial scale, field of view, acquisition rate, and dose budget. It may provide structural, molecular, chemical, metabolic, or mechanical information: OCT, quantitative phase imaging or holotomography, low-dose confocal or light-sheet microscopy, SRS, Brillouin microscopy, and AFM can each contribute in suitable implementations, although AFM is limited to accessible interfaces rather than internal volumes. Orthogonal validation should address what the backbone cannot identify. Agent penetration may be measured by tracer-assisted SRS or a functionally validated fluorescent conjugate; cellular response by reporter, viability, metabolic, or culture-based assays; matrix change by composition-specific probes or chemical imaging; and species identity by FISH, reporters, or spatial-omics endpoints. One implementation is a molecular identity channel registered to a repeated structural trajectory. This extends correlative microscopy to rugged 3D biofilm geometry, where registration is a central bottleneck (Krentzel et al. 2025). Cross-modal machine learning can help register or predict complementary channels; examples include virtual Gram staining (Işıl et al. 2025), mechanism-of-action classification from single-cell phase images (Sedaghat et al. 2025), species and spore discrimination from holographic images (Jo et al. 2017), and synthetic training data for biofilm segmentation (Holicheva et al. 2025)—but virtual staining does not replace independent validation because its outputs inherit the specificity, bias, and domain limits of the training labels. Direct multimodal acquisition should be organized around the biological question: agent-specific imaging plus a volumetric structural backbone for penetration; molecular cell-state channels plus longitudinal architecture for depth-resolved response; compositional imaging plus mechanical mapping for matrix remodeling; and a low-dose longitudinal backbone plus periodic identity and viability measurements for fate. Species identity can also be added by spatial-omics registration (Shi et al. 2020; Dar et al. 2021); however, no channel is ground truth by default. The reference measurement must itself be validated for the biological quantity claimed. Integration therefore requires a compatible cultivation format, quantitative calibration, cross-modality registration in a deforming geometry, and explicit dose and uncertainty reporting (Box 2).

**Conclusions and future directions**

3D imaging has changed what can be measured in biofilms, but no single modality provides an unambiguous account of structure and function across all relevant scales. The central comparison is therefore not label-based versus label-free imaging, nor is there a universally preferred platform. Modalities and implementations should instead be evaluated separately by contrast provenance, live volumetric capability, perturbation, demonstrated biofilm use, and the quantitative biological readout they support.

Intrinsic contrast does not automatically guarantee negligible perturbation, because illumination, mechanical contact, sample handling, and acquisition time can alter a living biofilm. Conversely, fluorescent, affinity-based, isotopic, and genetically encoded labels remain essential when the question requires species identity, molecular localization, viability, or pathway-specific information. The relevant criterion is empirical: whether the complete imaging implementation changes the process it is intended to report, and whether each indirect signal is validated against the biological quantity claimed.

4D biofilm measurement is therefore best treated as a validated measurement architecture rather than a two-class choice between labeled and label-free imaging. A longitudinal three-dimensional channel should be selected for low perturbation at the relevant scale and used to follow architecture, biomass-related contrast, molecular state, chemistry, or mechanics through growth or perturbation. Molecularly specific measurements may be acquired continuously when dose permits, or intermittently and at predefined validation points when continuity and specificity cannot be obtained in one channel. Depending on the biological readout, the backbone may be supplied by OCT, quantitative phase imaging or holotomography, low-dose light-sheet or confocal microscopy, SRS, or another validated platform; specificity may be supplied by fluorescence, Raman tracers, genetically encoded reporters, FISH, spatial omics, or terminal chemical and ultrastructural analysis.

Realizing this architecture requires more than placing two instruments beside one another. Cultivation formats must preserve the same biofilm across acquisitions; 3D datasets must remain registered despite growth and deformation; imaging dose and probe effects must be measured; and modality-specific signals must be reduced to parameters that can be compared across laboratories. If these conditions are met, 4D imaging can connect architecture with molecular identity, chemistry, mechanics, and fate while retaining continuity of observation. The next advance is unlikely to be a single winning modality, but a validated measurement architecture that combines longitudinal coverage with molecular specificity.

**Acknowledgements**

The authors thank Prof. Yilin Wu's laboratory, Department of Physics, The Chinese University of Hong Kong, for providing *P. aeruginosa* biofilm samples.

**Funding**

This work was supported by the National Research Foundation of Korea grant funded by the Korea government (MSIT) under Grants RS-2024-00442348 and RS-2024-00440577; the Korea Institute for Advancement of Technology (KIAT) through the International Cooperative R&D program under Grant P0028463; and the Commercialization Promotion Agency for R&D Outcomes (COMPA) funded by the Ministry of Science and ICT (MSIT) under Grants RS-2024-00440577 and RS-2026-25535856.

**Disclosure of interest**

Y.K.P. is cofounder and CEO of Tomocube Inc., which manufactures holotomography instrumentation. S.L. reports there are no competing interests to declare.

**Author contributions**

Y.K.P. conceived and framed the article. S.L. and Y.K.P. surveyed the literature, developed the modality framework, and wrote the manuscript. Both authors reviewed, edited, and approved the final version.

**Data availability**

This review generated no new large-scale datasets. The illustrative, previously unpublished HT time-lapse of a live *Pseudomonas aeruginosa* biofilm shown in Figure 3f is available from the corresponding author upon reasonable request. The panel is not used as evidence of comparative modality performance. All other data discussed are available in the cited primary publications.

Table 1

Table 1. Biofilm imaging modalities compared by contrast, live 3D capability, perturbation, demonstrated biofilm use, scale, and quantitative readout. The table does not rank modalities; each addresses a different combination of continuity, specificity, spatial scale, and quantitative interpretation. Demonstrated use refers to the closest biofilm evidence cited in this review and should not be interpreted as a modality-maturity score.

| Modality | Contrast provenance / primary signal | Live 3D status | Perturbation / specimen constraints | Closest demonstrated biofilm use | Scale / speed | Quantitative readout / integrated role |
|---|---|---|---|---|---|---|
| Confocal laser-scanning microscopy (CLSM) | Exogenous-label fluorescence; species, matrix, or cell-state channels | Demonstrated live volumetric | Probe/reporter dependence; depth attenuation, dose, bleaching | Live mature-biofilm architecture and antimicrobial-response imaging; endpoint meaning remains probe-dependent | Sub-µm lateral; typically ~50–100 µm depth; dose-limited 4D | Probe-defined intensity, segmentation, biovolume, matrix distribution, and cell trajectories; backbone or validation channel depending on dose and reporter stability |
| Light-sheet fluorescence microscopy (LSFM) | Exogenous-label fluorescence; reporter or dye | Demonstrated long-duration live volumetric | Reporter stability and model geometry; lower out-of-focus dose, not dose-free | Architecture, single-cell trajectories, collective flow, treatment response | Single-cell resolution; hours–days | Segmented cell positions, trajectories, architecture, and collective dynamics; molecularly specific backbone when cultivation and reporter stability permit |
| Multiphoton microscopy / FLIM | Protocol-dependent: exogenous multiphoton fluorescence or intrinsic autofluorescence lifetime | Live volumetric demonstrated in selected biofilm applications; capability is protocol-dependent | Photon dose, fluorophore or autofluorescence interpretation | Depth-extended fluorescence and lifetime-based metabolic or structural imaging in selected biofilms | Sub-µm lateral; greater optical depth than confocal; scan-rate dependent | Fluorescence intensity or lifetime distributions; biological interpretation requires fluorophore- or autofluorescence-specific validation |
| Optical coherence tomography (OCT) | Intrinsic optical backscatter | Demonstrated live volumetric; mesoscale | No contrast reagent; illumination dose; scattering and optical-access constraints | Engineered and clinical-device biofilm architecture, deformation, detachment | ~5–15 µm axial; mm depth/FOV; fast volumetric | Thickness, roughness, porosity, biovolume, deformation, and detachment; mesoscale structural backbone without intrinsic molecular identity |
| Quantitative phase imaging / holotomography (QPI/HT) | Intrinsic refractive index (RI) | Live 3D time-lapse demonstrated | No contrast reagent; illumination dose; depth/scattering limits | Single-cell response, early biofilm development, untreated live-biofilm dynamics | Submicrometre, diffraction-limited lateral resolution; reported depth ~50–100 µm; usable depth and volume rate are specimen- and system-dependent | Quantitative RI and RI-derived mass estimates under a composition-appropriate increment or calibration; identity and viability require orthogonal validation |
| Stimulated Raman scattering (SRS) | Endogenous or tracer-assisted molecular vibration; cited $D_2O$/alkyne cases are tracer-assisted | 3D depth stacks acquired in live specimens; repeated mature-biofilm volumes during treatment not established in the cited studies | Laser dose; isotope incorporation or tagged agent; parent equivalence requires validation | Agent localization and depth-resolved metabolic activity demonstrated in separate biofilm studies | Sub-µm lateral; reported depth ~50–100 µm; volume time depends on spectral channels, field of view, depth sampling, and dose | Molecular-vibrational intensity, agent distribution, or metabolic incorporation; chemical channel that can be paired with longitudinal structural imaging |
| Brillouin microscopy | Intrinsic acoustic-phonon frequency shift | Live 3D biofilm measurement demonstrated; repeated antimicrobial time series remain limited | Scan duration, photon dose, RI/density assumptions | Brillouin-shift gradients in live biofilms; challenged longitudinal integration not established | ~µm lateral and ~2 µm axial; point-scanning | Brillouin frequency shift as mechanical contrast; modulus interpretation requires RI, density, and model assumptions |
| In-liquid atomic force microscopy (AFM) | Intrinsic force–distance and topography | Live non-volumetric; surface/interface only | Mechanical probe contact and surface accessibility | Single-cell antimicrobial dynamics and nascent-biofilm or interface | nm topography; seconds at single-cell best case, minutes for biofilm field | Topography, force–distance response, and adhesion; interface-resolved mechanical companion, not an |

| Modality | Contrast provenance / primary signal | Live 3D status | Perturbation / specimen constraints | Closest demonstrated biofilm use | Scale / speed | Quantitative readout / integrated role |
|---|---|---|---|---|---|---|
| | | | | mechanics; no internal volumetric acquisition | | internal 3D validation method |
| Environmental / atmospheric SEM | Intrinsic electron scattering | Terminal or single-time-point hydrated-state snapshot | Beam, membrane-window, and vacuum constraints | Hydrated or uncoated biofilm ultrastructure | nm scale; beam-limited single time point | Channels, vesicles, eDNA, curli; terminal morphological reference |
| Soft X-ray tomography (SXT) | Intrinsic X-ray absorption | Terminal static 3D | Cryo/vitrified specimen; synchrotron; specimen-consuming | Cryo-vitrified biofilm three-dimensional structure | ~50 nm isotropic; single time point | Absorption/density contrast; 3D endpoint reference |
| SIMS / MALDI imaging | Analyte-specific mass-to-charge signal; SIMS uses sputter ionization and MALDI requires an exogenous matrix | Terminal chemical map | Terminal preparation and ionization; freezing, sectioning, sputtering, or matrix deposition depending on method | Biofilm molecular composition and agent gradients | Method-dependent; single time point | Molecular identity and distribution; terminal chemical validation |
| Conventional FIB-SEM | Stain-dependent electron-density contrast | Terminal static 3D | Fixation, dehydration, heavy metals, milling, vacuum | Mature-biofilm ultrastructure | Nanometre-scale serial-volume ultrastructure; resolution and volume depend on preparation and milling protocol | High-resolution ultrastructural endpoint |
| Cryo-ET with cryo-FIB | Intrinsic electron scattering | Terminal static 3D subvolume | Vitrification, cryo-FIB milling, high vacuum | Mature-biofilm subvolume and molecular machines | Sub-5 nm; single time point | Molecular/ultrastructural endpoint validation |

**Box 1 | Contrast provenance, perturbation, and validation are separate properties**

Contrast provenance describes where signal comes from; perturbation describes what the measurement changes; validation establishes whether the signal reports the claimed biology:

- Intrinsic contrast requires no contrast-generating reagent but may still involve illumination dose, mechanical contact, fixation, or destructive preparation.
- Tracer-assisted contrast uses an isotope, chemical tag, or radiotracer. Its chemical and biological effects must be measured rather than assumed negligible.
- Exogenous-label contrast supplies molecular, taxonomic, or functional specificity through dyes, affinity probes, viability stains, or genetically encoded reporters. Its effect depends on the probe and protocol.
- A reference channel is not ground truth by default. Reference status comes from validation against the biological endpoint being claimed.

For antimicrobial conjugates used as proxies for an unlabeled parent, activity and penetration should be compared directly. The goal is the least perturbative combination that still provides the identity required by the biological question.

**Box 2 | Requirements for integrated 4D biofilm measurement**

A longitudinal backbone and molecularly specific validation become informative only when the workflow preserves continuity and quantitative meaning:

- Compatible cultivation. Flow, substrate, medium, and sample holders must support repeated acquisition across the selected modalities.
- Cross-modality registration. 3D datasets must remain aligned despite biofilm growth, deformation, and detachment.
- Dose and probe controls. Illumination, contact, reporter stability, tracer incorporation, and tagged-agent equivalence must be measured for the specific experiment.
- Shared biological outputs. Thickness, biovolume, depth-resolved response, matrix change, and regrowth should be defined consistently, with modality-specific calibration and uncertainty reported separately.
- Linking structure to identity. Longitudinal structure and molecular or taxonomic identity are often supplied by different channels; their registration is a design requirement, not a failure of one modality.
- Scale bridging. Local microscopy, whole-animal burden, and clinical imaging require prespecified correspondence among endpoints because none validates another automatically.

Figures

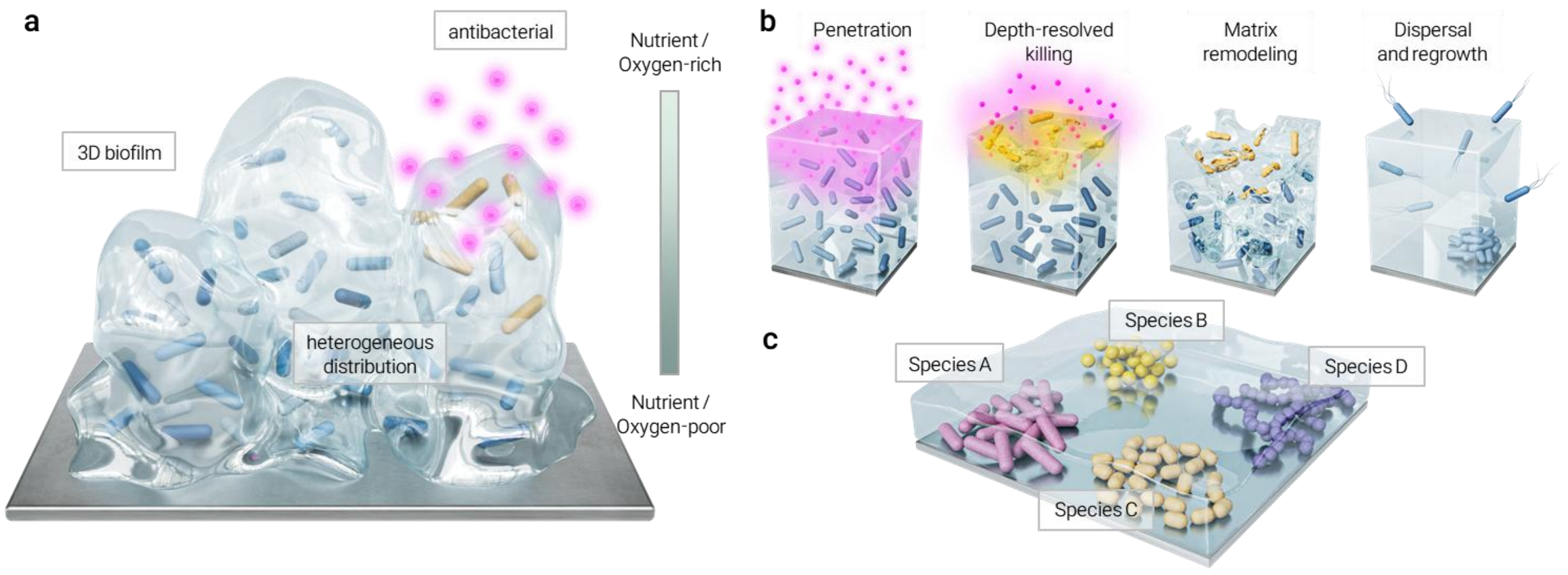

**Figure 1. Antimicrobial action as a 4D biofilm-imaging test case.** Three panels map biofilm biology onto the four structured questions. **(a)** A biofilm is an irreducibly 3D community whose cells occupy heterogeneous, depth-dependent states (shading) within an EPS matrix; an antimicrobial agent engages it locally rather than uniformly. Nutrient and oxygen availability declines with depth, establishing the gradients that stratify metabolic and tolerance phenotypes. **(b)** The post-treatment trajectory — penetration → depth-resolved killing → matrix remodeling → dispersal and regrowth — unfolds over time, and endpoint imaging cannot disambiguate its outcomes. **(c)** In polymicrobial biofilms, distinct species occupy spatially resolved niches. The depth, gradient, dynamics, and spatial structure in (a)–(c) are precisely what the planktonic MIC and 2D endpoint assays average away.

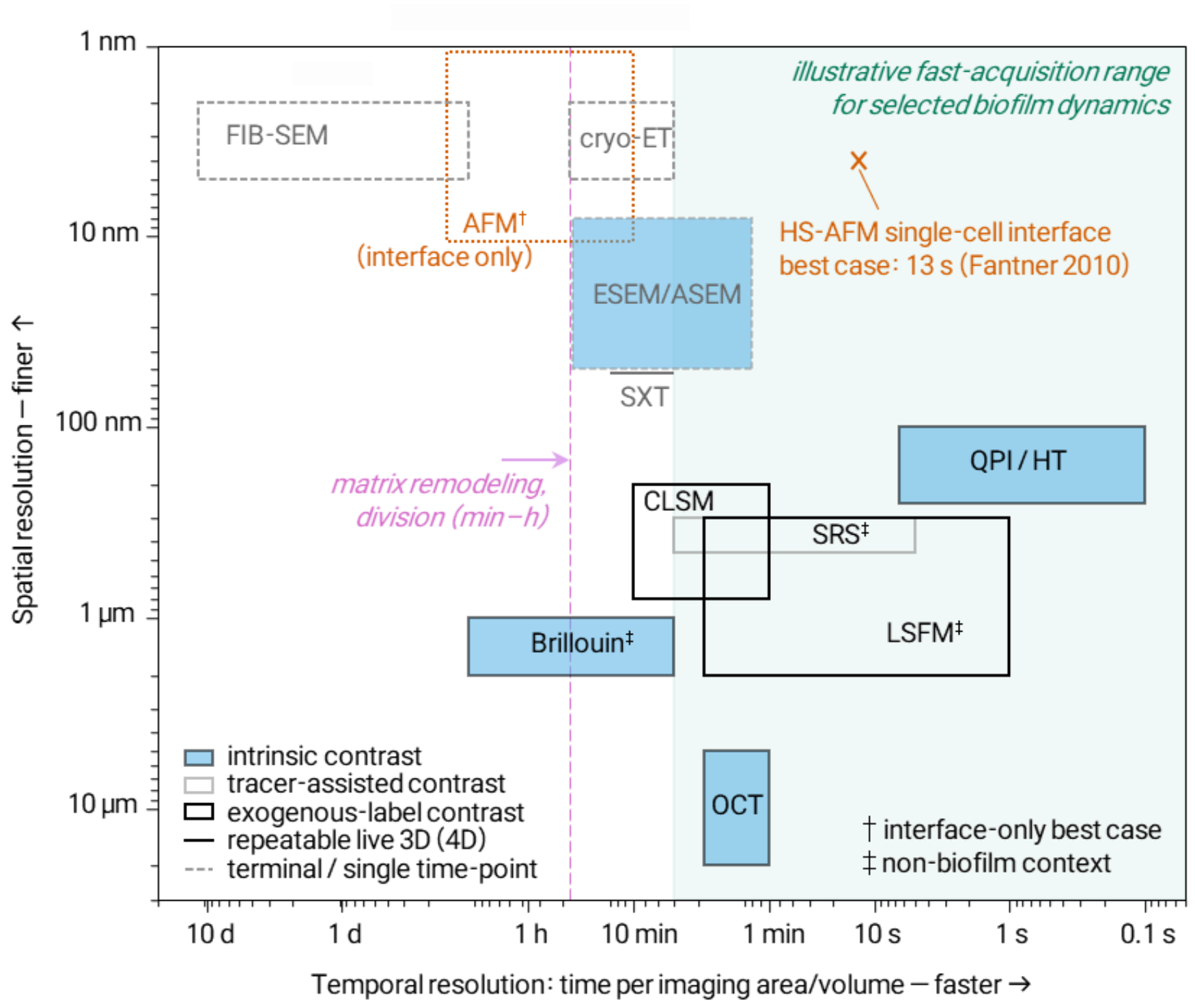

**Figure 2. Trade-offs in live 3D biofilm imaging.** Spatial resolution (vertical axis; finer upward) versus volumetric acquisition speed (horizontal axis; shorter time per 3D volume, hence faster, toward the right) for the principal imaging modalities applied to biofilms. Each modality is drawn as a box spanning its lateral-to-axial resolution and its fastest-to-slowest reported acquisition time; isotropic methods (e.g. soft X-ray tomography) collapse to a narrow band. Fill encodes contrast provenance using the figure legend: solid, intrinsic contrast; hatched, tracer-assisted chemical or isotopic contrast (e.g. stimulated Raman scattering with $D_2O$ or alkyne tags); open, exogenous-label. This encoding does not rank perturbation. Outline indicates nominal live 4D capability (solid) versus terminal or single-time-point acquisition (dashed); demonstrated biofilm use is reported independently in Table 1. Atomic force microscopy is shown as surface-only (it does not record an internal volume); the marked point is the single-cell high-speed-AFM best case (~13 s per image(Fantner et al. 2010)), whereas imaging a biofilm-relevant field is minutes per frame. The shaded region indicates acquisition rates compatible with selected biofilm dynamics at the field of view used in each source; it is not a preferred-modality zone. Modalities from different contrast classes can enter this regime under appropriate field-of-view and dose conditions. The plot is not a performance ranking. Acquisition times are not normalized to a common volume, and field of view, penetration depth, molecular specificity, dose, and quantitative validity are not represented on the two axes. Resolution values are from Table 1 and the references therein; reported fields of view differ by orders of magnitude. Times marked with a double dagger are for non-biofilm specimens of the same instrument class; ESEM/ASEM times are instrument-level values (Maruyama et al. 2012; Nishiyama et al. 2014; Sugimoto et al. 2016).

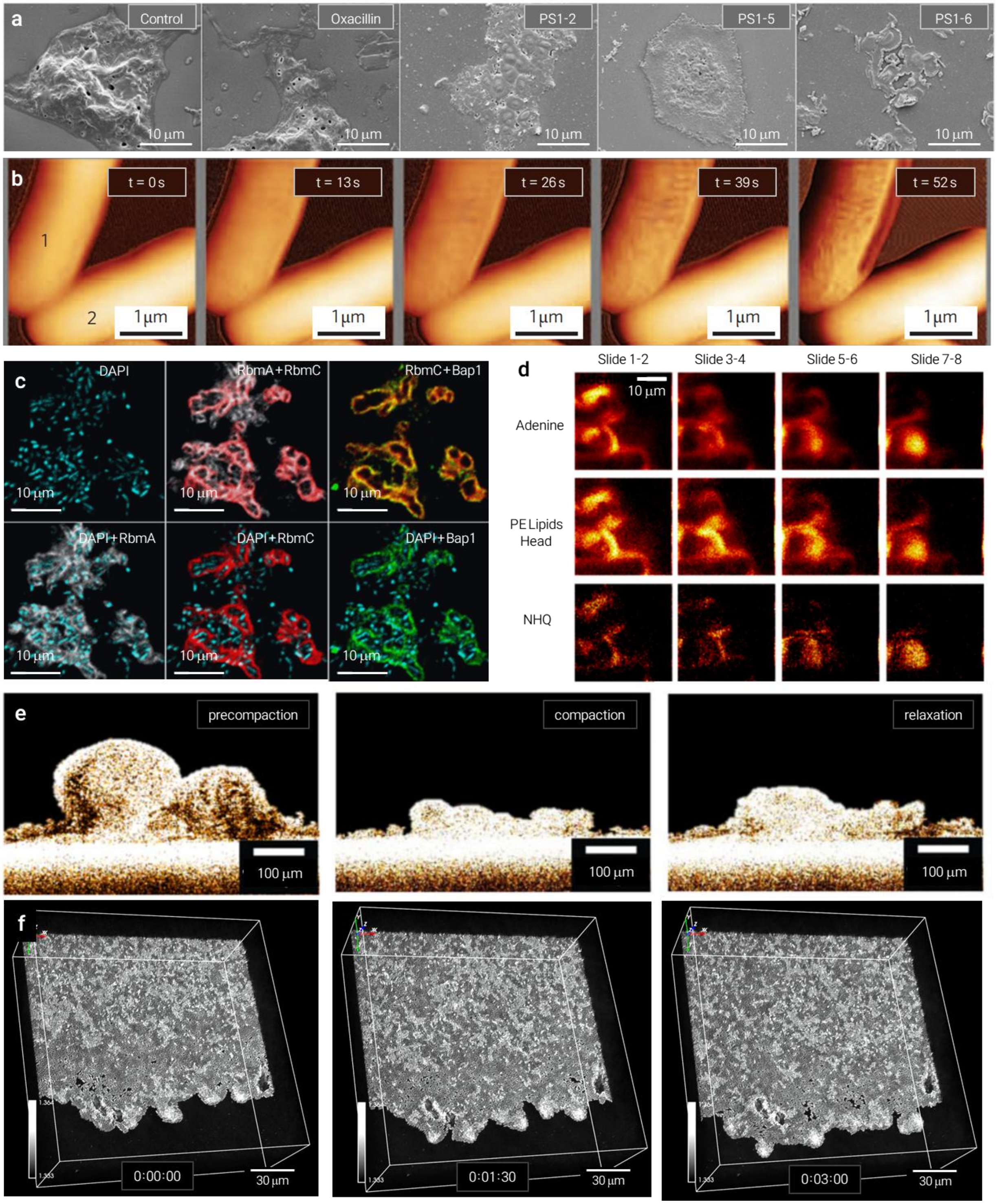

**Figure 3. Complementary structural, molecular, mechanical, and longitudinal imaging of biofilms.** Representative imaging modalities used to study bacterial biofilms across length scales and contrast mechanisms. **(a)** Scanning electron microscopy of preformed *Pseudomonas aeruginosa* DRPa 4007 biofilm, showing the disruptive effect of oxacillin and PS antimicrobial peptides (PS1-2/5/6) applied at their MICs for 24 h; scale bars 10 µm. Adapted with permission from Ref. (Park et al. 2019). **(b)** High-speed atomic force microscopy time series of *Escherichia coli* membrane disruption after exposure to the antimicrobial peptide CM15: sequential frames reveal rapid surface remodeling of one bacterium (1) while a neighbor (2) initially resists; scale bars 1 µm. Reproduced/adapted with permission from Ref. (Fantner et al. 2010). **(c)**

Confocal laser-scanning microscopy of *Vibrio cholerae* biofilm architecture: cells (DAPI) with extracellular-matrix proteins RbmA, RbmC, and Bap1 — RbmA localized around and within cell clusters, RbmC and Bap1 forming envelopes surrounding clusters; scale bars 10 µm. Adapted with permission from Ref. (Berk et al. 2012). **(d)** Orbitrap secondary-ion mass-spectrometry imaging of frozen-hydrated *P. aeruginosa* biofilm, mapping selected molecular species (adenine, phosphoethanolamine lipid headgroups, and a quinolone-related metabolite); scale bar 10 µm. Adapted with permission from Ref. (J. Zhang et al. 2020). **(e)** Optical coherence tomography of biofilm deformation under hydrodynamic perturbation — precompaction, compaction after increased permeate flux, and relaxation after restoring the original flux; scale bars 100 µm. Adapted with permission from Ref. (Valladares Linares et al. 2016). **(f)** Time-lapse 3D RI tomograms of a live *P. aeruginosa* biofilm acquired by holotomography, illustrating longitudinal 3D morphology and structural change in an untreated specimen; scale bars 30 µm. Panels (a)–(e) are reproduced or adapted from previously published studies with permission (to be obtained); panel (f) shows previously unpublished illustrative data and is not evidence of comparative modality performance.